\documentclass[12pt,NJP]{iopart}
\usepackage[english]{babel}
\usepackage{upgreek}
\usepackage{iopams}
\usepackage[dvips]{graphicx}
\usepackage{booktabs}
\usepackage[table]{xcolor}
\usepackage{tabularx}
\usepackage{colortbl}
\usepackage{graphicx}
\usepackage{caption}

\newcommand{\degree}{\ensuremath{^\circ}}

\newcommand{\iu}{{i\mkern1mu}}

\usepackage{color}
\definecolor{darkblue}{rgb}{0,0,0.5}
\definecolor{lila}{rgb}{0.3,0,0.3}
\definecolor{turq}{rgb}{0,0.1,0.4}
\definecolor{lightblue}{rgb}{0.7,0.7,0.9}
\usepackage{url} % that hyperlinks may be hyphenated correctly

\usepackage[pdftex,
 colorlinks=true,
 backref=page,
 linkcolor=darkblue, % usual links
 filecolor=red,
 citecolor=turq, % for bibliographic
 urlcolor=lila, % for Emails etc
 pdftitle={How to Build an Optical Filter with an Atomic Vapor Cell}
 pdfauthor={Denis Uhland, Helena Dillmann, Yijun Wang and Ilja Gerhardt},
 pdfsubject={},
 pdfkeywords={Atomic Vapors, Atomic Filters, Faraday Effect},
 breaklinks=false,
 plainpages=false,
 backref=false,
 bookmarks,
 bookmarksnumbered=true]{hyperref}

\begin{document}

\title{How to Build an Optical Filter with an Atomic Vapor Cell}

\author{Denis Uhland$^1$, Helena Dillmann$^1$, Yijun Wang$^1$, and Ilja Gerhardt$^1$}
\address{$^1$light \& matter Group, Institute for Solid State Physics, Leibniz University Hannover, Appelstrasse 2, D-30167 Hannover, Germany}
\ead{ilja.gerhardt@physics.uni-hannover.de}
\date{\today}

%% abstract
\begin{abstract}
The nature of atomic vapors, their natural alignment with interatomic transitions, and their ease of use make them highly suited for spectrally narrow-banded optical filters. Atomic filters come in two flavors: a filter based on the absorption of light by the Doppler broadened atomic vapor, i.e., a notch filter, and a bandpass filter based on the transmission of resonant light caused by the Faraday effect. The notch filter uses the absorption of resonant photons to filter out a small spectral band around the atomic transition. The off-resonant part of the spectrum is fully transmitted. Atomic vapors based on the Faraday effect allow for suppression of the detuned spectral fraction. Transmission of light originates from the magnetically induced rotation of linear polarized light close to an atomic resonance. This filter constellation allows selective acceptance of specific light frequencies. In this manuscript, we discuss these two types of filters and elucidate the specialties of atomic line filters. We also present a practical guide on building such filter setups from scratch and discuss an approach to achieve an almost perfect atomic spectrum backed by theoretical calculations.
\end{abstract}

\maketitle

%%%%%%%%%%%%%%%%%%%%%%%%%%%%%%%%%%%%%%%%%%%%%%%%%%%%%%%%%%%%%%%%%%%%%%%%%%%%%%%%%%%%%%%%%%%%%
%% Intro
%%%%%%%%%%%%%%%%%%%%%%%%%%%%%%%%%%%%%%%%%%%%%%%%%%%%%%%%%%%%%%%%%%%%%%%%%%%%%%%%%%%%%%%%%%%%%

\section{Introduction}
%% Intro into the field (what can hot atomic vapors do for us?)
One of the first experiments in atomic physics was the observation of atomized alkali salts in flames~\cite{kirchhoff_1860}. Over time, evacuated glass cylinders filled with a small amount of an atomic element such as sodium, potassium, or rubidium replaced those open flame experiments. The advantages of vapor cells are their robustness, the freedom to manufacture them in almost arbitrary geometries, the flexible choice of the atomic species, and their convenient handling. All this makes them a great tool for a wide range of applications. To name a few, they can be put into use as magnetic~\cite{budker_nature_2007, groeger_epjd_2006, savukov_prl_2005} and electric field sensors~\cite{sedlacek_prl_2013}, as storage media in quantum optics~\cite{phillips_prl_2001,siyushev_n_2014}, and also as atomic line filters~\cite{ohman_soa_1956, yeh_ao_1982, dick_ol_1991}, which is the topic of this tutorial article.

%% Zeeman
Experiments have shown that atomic vapors effectively block light on specific lines~\cite{kirchhoff_1860}. Such effects occur when optical resonant light gets absorbed by the atom. In contrast to solids or liquids, atoms in their vapor phase act as an absorbing medium and do not experience the same spectral broadening. The broadening is usually limited by the atom's velocity distribution, also known as Doppler broadening. The underlying Maxwell-Boltzman velocity distribution ranges from a few hundred to a few thousand meters per second. The resulting GHz wide bandwidth is at least two orders of magnitude narrower than good commercial dichroic filters. Subsequently, vapor cells present a useful instrument to block unwanted spectral ranges. These ``Doppler filters'' act as narrow-band notch filters, which only block the light near and on an atomic transition, but are transparent to all other spectral components, limited solely by the residual absorption of the vapor cell's windows.

%% Righi filter
The optical properties of atomic vapor cells change under the influence of an external magnetic field. The Zeeman effect~\cite{zeeman_nature_1897} splits the atomic energy structure into sub-levels with different optical transition frequencies. Augusto Righi used this effect and observed that individual polarization components of the light are differently affected by the magnetically shifted sub-levels~\cite{righi_cr_1897, righi_cr_1899}. This narrow-band notch filter became known as the ``Righi-filter''. Since it compares well to the other filters described in this manuscript, we will not dive into any detailed discussions.

%% Macaluso-Corbino
In 1898, Macalouso and Corbino studied polarized light from the sun. In their experiments, the light passed a flame that included some amount of sodium. A second linear polarizer placed behind the sodium flame suppressed (``crossed-out'') the light effectively. They realized that the light overcame the blockade of the second polarizer when a magnetic field was introduced~\cite{macaluso_inc_1898, budker_rmp_2002}. That was only possible if the linear polarized light performed a rotation. This effect became known as the ``Macaluso-Corbino effect''. In the 1950s, this effect allowed Yngve \"{O}hman to implement a narrow-band atomic filter~\cite{ohman_soa_1956}, surpassing the quality figures of the earlier developed Lyot-filters~\cite{lyot_1933}, which was commonly used for astronomic observations~\footnote{Yngve \"Ohmann re-invented the Lyot-filter, which was first discovered by Bernard Lyot~\cite{wyller_1990}}. In an independent report later, Kohler and Novick linked the Macaluso-Corbino filter~\cite{kohler_qr_1962} to the Faraday effect. Unlike the Macaluso-Corbino effect, which describes the rotation of the polarization in the vicinity of an atomic transition (resonant), the Faraday effect is not necessarily based on the presence of atomic resonances but rather describes the effective rotation due to a longitudinal magnetic field in any medium. The Faraday effect can also appear without external magnetic fields as long as the interaction induces some magnetic effects on, e.g., the surface of a ferromagnet. Due to the more general definition, the term ``Faraday filter'' gained popularity. Those Faraday filters suppress all frequencies of light except the ones in proximity to an optical atomic transition. Therefore, such filters are suited for GHz-wide band-pass filters. Research on Faraday filters became more popular in the 1970s and 1980s~\cite{agnelli_sp_1975, roberts_jpb_1980, yeh_ao_1982, barkov_oc_1989}. Later, novel names such as the ``Faraday anomalous dispersion optical filter'' (FADOF) were coined~\cite{dick_ol_1991, yin_procieee_1992}.

%% Voigt filter
For any atomic filter arrangement, the orientation of the external magnetic field can be arbitrary. Common choices are a \textit{longitudinal} magnetic field along the laser axis or a \textit{transverse}  magnetic field, where the magnetic field is orthogonal to the propagation of the light. In both cases, the excitation light experiences resonant birefringence effects~\cite{halpern_pr_1964}. A \textit{transverse} magnetic field leads to the Voigt effect, discovered by Woldemar Voigt in 1887~\cite{voigt_1887}. This yields linear birefringence, which makes the vapor act like a resonant lambda half waveplate. Filters based on the Voigt effect were realized by using alkali metals in vapor cells like cesium~\cite{menders_ol_1992} or rubidium~\cite{yin_oe_2016, keaveney_jpb_2019}. On the other hand, a \textit{longitudinal} magnetic field causes circular birefringence, which rotates the polarization of the light in the vicinity of an atomic resonance. This manuscript focuses on circular birefringence and shows how vapor cells can act as filters when utilizing the Faraday effect for optical filtering.

%% History of atomic vapor filters
Atomic filters have a wide range of applications. Starting from astronomical, mostly solar, observations~\cite{ohman_soa_1956, cimino_sp_1968}, the field went over laser locking applications~\cite{sorokin_apl_1969} to optical communication~\cite{bomke_ao_1977}. In a chain of experiments, filters with different types of metals were under study like absorptive sodium filters~\cite{cimino_ao_1968}, filters around the sodium doublett~\cite{agnelli_sp_1975}, or filters based on rubidium~\cite{dick_ol_1991}, mercury~\cite{fork_ao_1964}, potassium~\cite{yin_oc_1992} and cesium~\cite{chen_jpb_1987, chen_oc_1990}. Several non-alkali atoms like bismuth~\cite{roberts_jpb_1980} or samarium~\cite{barkov_oc_1989} also found their way into the vapor cell for filtration purposes. Research attempts on excited atomic state Faraday filters were also reported~\cite{billmers_ol_1995, zhang_oc_1998,popescu_phd_2010, rudolf_oe_2014}, partly utilizing so-called ``see-through hollow cathode lamps''.

%% to date applications
To date, the range of applications extends to fields where atomic vapor cells perform light imaging detection and ranging (LIDAR)~\cite{cheng_fooic_2008, yong_ol_2011, huang_ol_2009, fricke-begemann_ol_2002} and terrestrial seawater observations by the Brillouin-lidar project~\cite{popescu_phd_2010, rudolf_oe_2014}. Groups like the Solar Activity Magnetic Monitor (SAMM and SAMNet) were founded for space weather research and forecast and developed a flare warning system based on vapor filters~\cite{erdelyi_swsc_2022}. Filter combinations are also possible, which can eliminate different spectral bands~\cite{beckers_ao_1970}. A model for maximizing the transmission with arbitrary magnetic field directions was recently reported~\cite{rotondaro_josab_2015}. All alkali atoms except atomic lithium were used as atomic line filters. Studies of magneto-rotational effects in lithium were so far only performed with cold atomic ensembles~\cite{franke-arnold_jpb_2001}.

%% this paper is organized as follows*
This manuscript starts with an introduction to Doppler filters. We first discuss the basic principles for absorptive filters and show the impact of external parameters like the temperature to tweak the filtration ability of the vapor cell. We then introduce a longitudinal external magnetic field which leads to the Faraday effect. We then show a theoretical analysis of the Faraday effect for sodium vapor and discuss the fundamentals of an atomic Faraday filter. The reason why the theory part revolves around sodium vapor is because of its descriptive spectrum. In the last part, we apply those theoretical methods and show an exemplary setup of a Doppler and a Faraday filter based on rubidium. Rubidium is the medium of choice due to its experimental significance.

\begin{figure*}[t]
	\includegraphics[width=\textwidth]{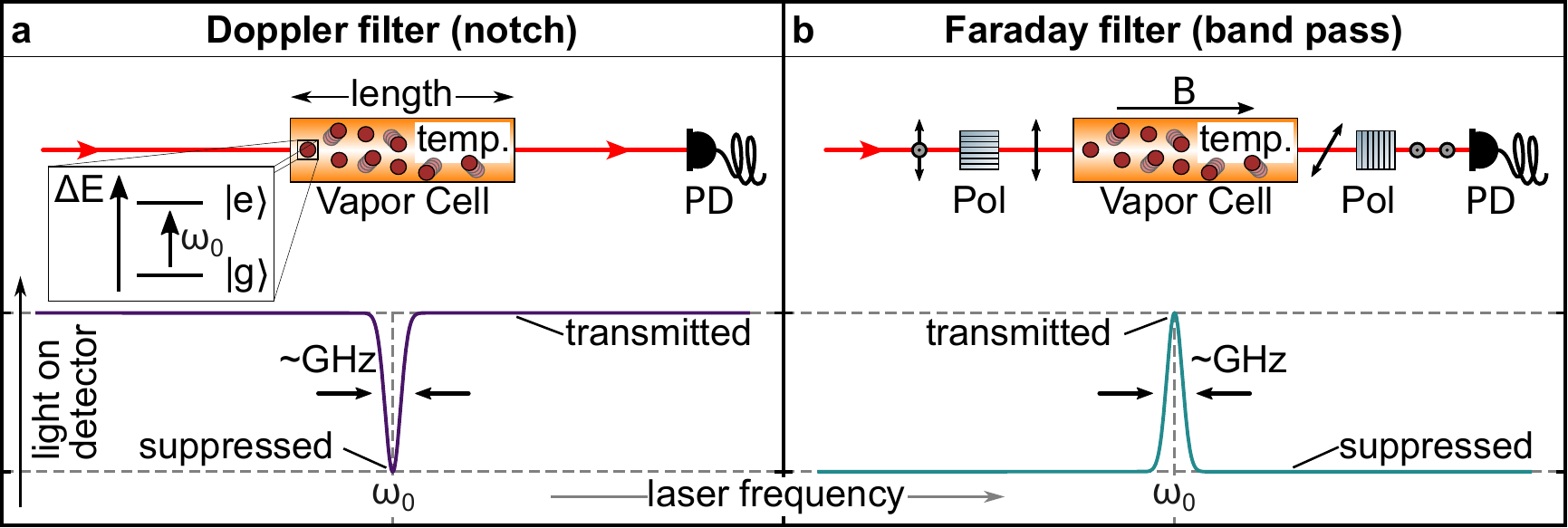}
	\caption{\textbf{Two types of atomic filters.} The Doppler filter absorbs light on and near an atomic transition, and acts as a notch filter. The Faraday filter suppresses all light except a small range around the atomic transitions and acts as a bandpass filter. \textbf{a)} Atomic notch filter with an atomic vapor cell with length $L$ and at a temperature $T$. The scanning range of the laser is within a range of several tenths of GHz across the atomic resonance. For far-detuned laser light, the atomic vapor is transparent. \textbf{b)} In the Faraday filter, the atomic vapor cell is placed between two crossed polarizers. An applied longitudinal magnetic field $B$ rotates linear polarized light by the atomic vapor (arrows). Thereby, the light can traverse both crossed polarizers. Resonant light gets transmitted while far detuned light is suppressed by the polarizers.} 
	\label{fig:01}
\end{figure*}

\section{The Absorptive Filter}

%% The basic explanations, Fig. 1a
This section focuses on the Doppler filter and discusses how external parameters affect the vapor's filtering ability. To set up the Doppler filter, one only needs to put a hot vapor cell in the pathway of the laser beam. The nature of atoms and their ability to absorb light allows hot atomic vapors to act like a notch filter for specific frequencies. It suppresses spectral components around an atomic transition while transmitting far-detuned frequencies. Figure~\ref{fig:01}a) shows the fundamental absorption for a single atom represented by the two-level system.

%% Absorption and hyperfine structure
The filtration effect then results from the transfer of atoms from the ground state $|g\rangle$ to the excited state $|e\rangle$. The energy states result due to the interplay of the momentum of the sodium valence electron and nuclear spin momentum, known as the hyperfine structure. The quantum numbers for the sodium ground state $|g\rangle$ are $l = 0$ and $s = 1/2$ for the electron (j = l + s), and $i = 3/2$ for the nucleus. This adds to $f = i \pm j = 3/2 \pm 1/2 = 1$ and $2$ for the whole atom. The same calculation applies for the excited state $|e \rangle$, where $l=1$ leads to $f = 0, 1, 2, 3$ for the sodium D$_2$ line. The excited atoms relax over time and emit light at all solid angles. A spatially limited detection in the forward direction ensures that this spontaneous emission does not significantly affect the detection of the atomic spectrum.

%% Doppler temperature and broadening
Several variables influence the role of the atoms as an active optical filter medium, like the choice of the atomic species, the length of the vapor cell $L$, and the temperature $T$. A temperature change affects the number of atoms in the vapor phase in a defined volume and their temperature-dependent velocity. With the temperature, the number of atoms in the vapor phase of the active medium changes in an almost exponential fashion~\cite{steck_sodium_numbers} and follows the Antoine equation~\cite{antoine_1888}. Therefore, it is possible to change the number density of accessible atoms by several orders of magnitude when changing the temperature only a bit. We will see below that this change in the density immediately affects the absorbance of the medium. At the same time, the average velocity of the atoms increases, which leads to a broadened absorption spectrum. This so-called Doppler broadening of the atomic vapor is a dominant feature for the edge-steepness of the absorption feature, given by the Maxwell-Boltzmann distribution $f(v) ~ $d$^3v =\left(\frac{m}{2 \pi kT}\right)^{3/2} \, \exp \left( -\frac{mv^2}{2kT} \right) ~ $d$^3v$.

%% Beer-Lambert
Following up on the Doppler broadened absorption feature, we can describe the intensity loss of the light which passes through the vapor cell by Beer-Lambert's law. When the incident light in front of the cell shows an intensity of $I_{\rm{in}}$, the light intensity right after the vapor cell drops to $I_{\rm{out}}$. The laser intensity reduces exponentially while traversing the cell due to the interaction with the atomic vapor. Each atom in the optical beam path can be associated with an atom-specific and wavelength-dependent extinction cross-section $\sigma$. The number of accessible atoms in the optical beam path is given by the atomic density $\rho$ and the optical path length $d$. Beer-Lambert's law then reads:

\begin{equation}
	log_{10} \left( \frac{I_{\rm out}(d)}{I_{\rm in}(0)} \right) = -\rho \cdot \sigma \cdot d 
	\label{eq:beer}
\end{equation}

\noindent
%% optical density
We represent the exponential suppression of light in the course of the cell with the decadic intensity ratio (effective transmission), commonly known as the ``optical density'' $OD$.

\begin{equation}
	log_{10} \left( \frac{I_{\rm in}(0)}{I_{\rm out}(d)} \right) = OD
\end{equation}

\begin{figure*}[t]
        \includegraphics[width=\textwidth]{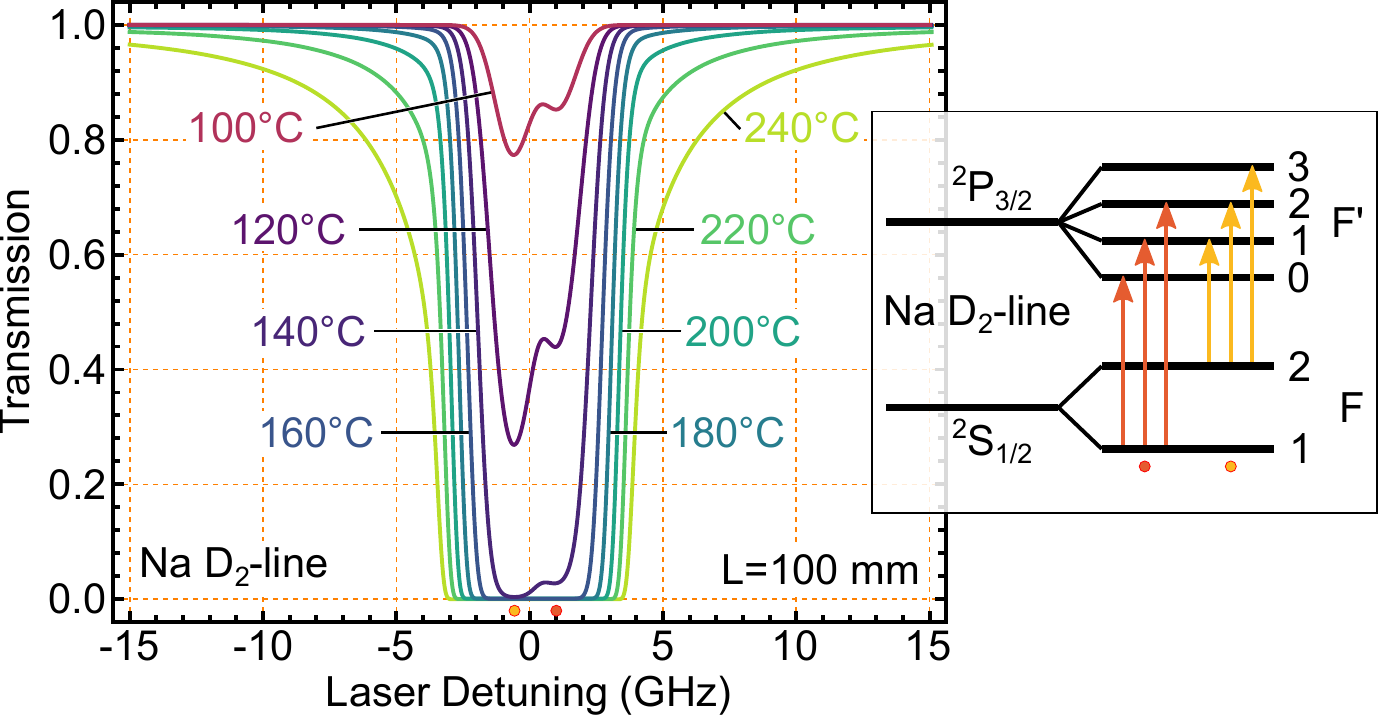}
	\caption{\textbf{Temperature dependency of the sodium absorption features around the D$_2$ transition.} The dots at the bottom denote the frequency of the hyperfine transitions. All six hyperfine transitions are visualized and color coded. The splitting of these six transitions is in the MHz range and is barely distinguishable by the dots. Due to the typical Gaussian lineshape caused by the Doppler broadening, the line width increased due to the linear scale. At elevated temperatures, the wings of the spectrum are dominated by the natural line width of the intrinsic atomic transition.}
        \label{fig:2}
\end{figure*}

\noindent
%% Fig 2 (example: sodium, temp 100-220degC), edge steepness
Figure~\ref{fig:2} shows examples of the sodium absorption features for different vapor temperatures and fixed cell length (100~mm) around the sodium D$_2$ line ($\approx$ 589.0~nm). The spectrum shows two major absorption lines. Each of the two broadened absorption features includes three transitions around the D$_2$ line, which can not be resolved for high temperatures (splitting is below 100~MHz). For low temperatures (100-140\degree C), the sodium ground state splitting (hyperfine splitting) is $\Delta \nu \approx 1.77$~GHz apart. Those features will merge for higher temperatures. The dots at the bottom of the plot indicate the positions of the hyperfine transition frequencies. The corresponding edge-steepness (10-90\%) is around 875~MHz for atomic sodium at 150\degree C as given by the Doppler broadening of the sodium vapor.

%% combination -> double exponential
Along with the increase in temperature and the velocity of the atoms, the vapor pressure also increases. The vapor pressure and the suppression of the transmitted light follow exponential laws. Minor temperature changes impact the vapor's transmittance due to this ''double-exponential'' temperature dependence significantly. For example, the optical density of a sodium vapor cell at 100~\degree C is less than 0.3, at 150~\degree C is 11.3, and at 200~\degree C is 215.7 Another increase in temperature by just 5~\degree C, to 205~\degree C, increases the decadic absorbance to 279.6.

%% wings go down
At temperatures below 200~\degree C, the spectral width of the sodium absorption line is dominated by the Doppler broadening, following a Gaussian profile. Any increase in temperature ``pulls'' the spectrum down. Therefore, in a linear representation, the absorption spectrum gets broader, with virtually no change in the edge steepness. This intervenes with the natural linewidth of the atomic transition, which underlies a Lorentzian shape. For higher temperatures, e.g. above 200~\degree C for sodium, the absorbance in the cell is so high that the underlying fundamental Lorentzian transition of sodium starts to alter the wings of the spectrum exceeding the Doppler broadening. Those features scale proportional to $1/\Delta^2$, where $\Delta$ is the detuning. The lineshape is the convolution of the Gaussian and Lorenzian, known as the Voigt line profile.

%% Applications, extended
Atomic filters in the Doppler configuration have been part of numerous research projects on light filtration. An early achievement was the filtration of light using spectral lamps~\cite{bender_prl_1958}. Here, the authors used atomic vapor cells to filter the spectral components for the two isotopes of naturally abundant rubidium, $^{85}$Rb and $^{87}$Rb. The approach was to set up a vapor cell with the isotope $^{85}$Rb, which suppressed the spectral response of this isotope, allowing the observation of the isotope $^{87}$Rb. The development of comparable methods allowed the usage of vapor cells for light imaging detection and ranging (LIDAR). The establishment of such filters with barium and other systems began in the early 1980s~\cite{shimizu_1983} . Other applications, like Raman spectroscopy, desired narrow-band rejection filters, such that low Raman shifts were accessible~\cite{indralingam_analchem_1991, pelletier_applspec_2003}. In microscopy, atomic Doppler filters allow the detection of the luminescence of DNA strains by suppressing the excitation laser light and even enhancing the photon detection efficiency by 15\% against the best available commercial filter~\cite{uhland_epj_2015}.

%% outlook
This summary of the Doppler filter and absorption lines should have brought the general concept closer to the reader. In the following section of this tutorial, we manipulate the Doppler spectrum by external magnetic influences, which inevitably introduces the Faraday filter.

\section{The Faraday Filter}
%%% Faraday filter

%% basic explanations, Fig. 1b
As for the Doppler filter, the central element of the Faraday filter is a hot vapor cell. Figure~\ref{fig:01}b shows the fundamental principle of a Faraday filter. Two linear polarizers sandwich the vapor cell. The polarization axes are aligned orthogonal to each other. This configuration ensures the suppression of the linear polarized light, which gets blocked by the second polarizer. If a longitudinal magnetic field is applied, it rotates the linear polarized light when it travels through the atomic vapor. This effect has a narrow-banded nature, which entitles atomic vapor filters as competitive candidates for bandpass filters.

%% Theoretical explanation of rotation
\subsection{Theory of the Faraday Filter - from Absorption to Rotation}
We suppose the light with frequency $\omega$ travels along the z-axis. That means the polarization of the light is in the $x$ and $y$ plane, which means for the initial wave:

\begin{equation}
	\vec{E}(z, \omega) = E_{0,x} ~\exp[i (\omega t - kz)] \hat{e}_x + E_{0,y} ~\exp[i (\omega t - kz )] \hat{e}_y
        \label{eq:initialwave}
\end{equation}

\noindent
Now, the light passes the atomic vapor. The response of the absorbing atomic vapor to the light is known as susceptibility. Resonant light gets absorbed by the optically active medium. We should therefore define the complex susceptibility $\chi_{\pm} = \chi_{\pm}^\prime + \iu \chi_{\pm}^{\prime \prime}$. The complex susceptibility acts on each circular component of the linear polarized light individually. We can immediately link the response of the atomic vapor to the refractive index with $n = \sqrt{1 + \chi}$. The refractive index also appears in the expression of the wavevector, which reads $| \vec{k} | = n \omega /c$. Similar to the susceptibility, the expression of the refractive index consists of a real and imaginary part that acts on the different circular parts individually, $n_{\pm} = \nu_{\pm} + \iu/2 \kappa_{\pm}$. We can modify the expression of the incident light (equation~\ref{eq:initialwave}) to:

\begin{equation}
	\vec{E}(z, \omega) = E_{0,x} ~ \exp \left[ i \left( \omega t - \frac{n_x \omega}{c}z \right) \right] \hat{e}_x + E_{0,y} ~ \exp \left[i \left( \omega t - \frac{n_y \omega}{c}z \right) \right] \hat{e}_{y}
        \label{eq:initialwave2}
\end{equation}

\noindent
Consider the beginning of the cell at $z=0$, which defines the incoming wave $\vec{E}_{in}$ as:

\begin{equation}
	\vec{E}_{in}(0, \omega) = E_{0,x} ~ \exp \left[ i \omega t \right] \hat{e}_{x} + E_{0,y} ~ \exp \left[ i \omega t \right] \hat{e}_{y}
        \label{eq:in}
\end{equation}

\noindent
If we align the first polarizer into the x-plane, equation~\ref{eq:in} simplifies to:

\begin{equation}
	\vec{E}_{in}(0, \omega) = E_{0,x} ~ \exp \left[ i \omega t \right] \hat{e}_{x}
        \label{eq:inx}
\end{equation}

\noindent
The same idea applies to the light, which leaves the vapor cell after length $L$. Despite the polarization before the cell, we do have to consider that a rotation occurred. So the polarization of the light after the vapor cell $\vec{E}_{out}(L, \omega)$ projects again somewhere to the $x$ and $y$ plane. Since it is still a planar wave, it shows the same form as in equation~\ref{eq:initialwave2} but for $z=L$. Furthermore, we can express the field by its circular components, $\hat{e}_x =-1/\sqrt{2} \left( \hat{e}^{+} - \hat{e}^{-} \right)$ and $\hat{e}_y = \iu/\sqrt{2} \left( \hat{e}^{+} + \hat{e}^{-} \right)$. In the end, we measure the effective transmission after the second polarizer, which has to be aligned in the $y$ direction. Therefore, we project $\vec{E}_{out}$ to $\hat{e}_y$. We use the identity $e^{-\iu x} + e^{+\iu x} = 2 \cos{x}$, the sum of the the absorption coefficients $\Delta \kappa = 1/2 (\kappa_{+} + \kappa_{-})$ and the sum of the refractive indeices $\Delta n = n_{+} + n_{-}$ for the circular light components. The expression of the effective transmission after the second polarizer is then:

\begin{equation}
	T = \left| \frac{\vec{E}_{out} \hat{e}_y}{\vec{E}_{in}} \right|^{2} = \frac{1}{4} 
        \Biggl(
        \underbrace{e^{- \kappa_{+} L} + e^{-\kappa_{-} L}}_{\mbox{Righi effect}} 
        -
        \underbrace{ 2 \cos \left( \frac{\omega}{c} \Delta n L \right) 
        \cdot 
        e^{-\Delta \kappa L}}_{\mbox{Macaluso-Corbino effect}} \Biggr)
        \label{eq:transmission}
\end{equation}

%\noindent
%In the last step, we use the identity $e^{-\iu x} + e^{+\iu x} = 2 \cos{x}$ and $\Delta \kappa = 1/2 (\kappa_{+} + \kappa_{-})$ and $\Delta n = n_{+} + n_{-}$. 

\noindent
This result shows two terms, the Righi and the Macalouso-Corbino effect. The Righi effect describes the absorption ($\kappa_{\pm}$) of the circular light components by the Zeeman levels. The Marculoso-Corbino effect, however, includes an angle in the cosine term, which is the rotation angle of the polarized light for a given frequency $\omega$.

\begin{equation}
        \Phi(\omega) = \frac{\omega}{c} \Delta n L
        \label{eq:rot}
\end{equation}

\noindent
The difference of the refractive indices on the circular components is proportional to the rotation angle $\Phi$. The circular light components of the light interact differently with the atomic vapor. This leads to an overall shift resulting in a net rotation of the polarization. Figure~\ref{fig:03} visualizes this. The calculation above follows~\cite{harrell_josab_2009} and is accessible as an open-source program called ElecSus~\cite{zentile_cpc_2015, keaveney_cpc_2018}.

%% ENBW
The peak transmission, $T_{\rm max}$, is one of the crucial parameters for optimizing the filter. The simple optimization of the peak transmission often also increases the side wings of the filter. At the same time, the rejection ratio for such unwanted spectral components gets reduced. Subsequently, the ``equivalent noise bandwidth'' (ENBW) was termed~\cite{dick_ol_1991}, which describes the achievable signal-to-noise ratio for white incident light. The ENBW is defined as:

\begin{equation} 
	\textrm{ENBW} = \frac{1}{T_{\rm{max}}} \int_{-\infty}^{\infty}{T(\omega) \rm{d} \omega} 
\end{equation}

%% FOM
\noindent
However, this is not sufficient for the optimal operation of the filter. The ENBW can be optimized, while the maximum transmission is on small and negligible values. For that, an additional optimization parameter has been established~\cite{kiefer_srep_2014}. The so-called ``figure of merit'' (FOM) maximizes the peak transmission while reducing the ENBW and can be calculated by taking the ratio of the peak transmission and ENBW. The higher this value is, the better the overall performance of the filter. Maximizing this value is equivalent to the best operation point for the filter. 

\begin{equation}
	\mbox{FOM}=T_{\rm max}/\mbox{ENBW} 
\end{equation}

\subsection{The Impact of the Magnetic field - the Zeeman Effect}
%% Zeeman
Filters in the Faraday configuration use a magnetic field $\vec{B}$ parallel to the wavevector $\vec{k}$ of the excitation light. The latter defines the quantization axis. With the interaction Hamiltonian, we can calculate how an external magnetic field splits and shifts the atomic energy states.

\begin{equation}
	H = H_{\rm{0}} + H_{\rm{HFS}} + H_{\rm{Zeeman}}
\end{equation}

\noindent
The Hamiltonian $H_{0}$ describes the unperturbed case and consists of a kinetic and a potential term. The hyperfine Hamiltonian ($H_{\rm{HFS}}$) describes the interaction between the momenta of the nucleus and the electron $\vec{i}$ and $\vec{j}$ times a coupling constant $a_{hf}$. Also, consider that external influences, like a magnetic field, can decouple the spin interaction. In such a case, each spin contribution of the atom, which carries a magnetic moment, couples individually to the magnetic field, denoted by the magnetic quantum number $m$. The multiplicity of the splitting of $f$ states scales like $2f+1$. An example of the sodium ground state would be the $F=1$ state, which splits into three sub-states with the magnetic quantum numbers $m_f = -1, 0, 1$, known as the Zeeman splitting, illustrated in figure~\ref{fig:03} f). Summarized, the interaction between the nucleus, represented by $i$, and the electron, represented by $j$, and finally, the projection of the momenta to a magnetic field defines the interaction Hamiltonian:

\begin{equation}
        H_{int} = H_{HFS} + H_{Zeeman} = \frac{a_{hf}}{\hbar^2} \vec{i} \cdot \vec{j} + \mu_{B} \vec{B} \cdot \frac{g_{s}\vec{s} + g_{l}\vec{l} + g_{i}\vec{i}}{\hbar}
\end{equation}

\noindent
Here, $g$ represents the Land\'e factors of the spins, and $\mu_{B}$ is the Bohr magneton. Each of the Zeeman levels experiences a shift proportional to the magnetic field $\propto \mu_{B}m_{f}g_{f}B$. The transition between these hyperfine states follows momentum conservation and shows a dependence on the polarization of the excitation light. If the light shows a right-handed circular polarization, the electron's momentum gets ''kicked'' by the absorbed photon, which increases the magnetic quantum number from $m_{f} \rightarrow m_{f} + 1$. The same idea applies to left circular polarized light, which would decrease the magnetic quantum number $m_{f} \rightarrow m_{f} - 1$. Linear polarized light, which oscillates perpendicular to the quantization axis, would not be absorbed. Therefore, the Zeeman splitting of the absorption Doppler spectrum only shows an impact on the circular transitions for longitudinal magnetic fields.

\subsection{Working Principle of the Faraday Filter}
%\noindent
%% Wing and center operation

%% Wing and center operation
The example of a sodium Faraday filter is very suitable for a theoretical discussion and explanation of the fundamental working principle of a Faraday filter. The Doppler broadened absorption line splits into the two Zeeman components. When the field is sufficiently large, it opens a window in which the absorption is low, but the optical rotation simultaneously amounts to 90\degree. Therefore, no Righi effect limits the transition of the filter. Therefore, the spectrum appears as a peak on a pedestal. Furthermore, the left-most and right-most side displays a higher transmission than the 50\% limited Righi-type side-plateaus. Due to the large splitting and the opened transparancy window, where no Doppler absorption occurs, the transmission in the center exceeds the side wings. This operation is called the center operation of the Faraday filter. 

%% side operation
For higher temperatures and low magnetic fields, the rims of the Doppler absorption spectrum exhibit strong optical rotation. Then the so-called wing operation of a Faraday filter occurs. It is evident, that in this case the anomalous dispersion of the Faraday filter does not show any relevance for most of the vapor filters. In the past, the misconception that anomalous dispersion are the dominant effects of those filters found approval, testified by the name ``FADOF''. Nowadays, this concept is disproved~\cite{gerhardt_ol_2018}.

\begin{figure*}[t]
	\includegraphics[width=\textwidth]{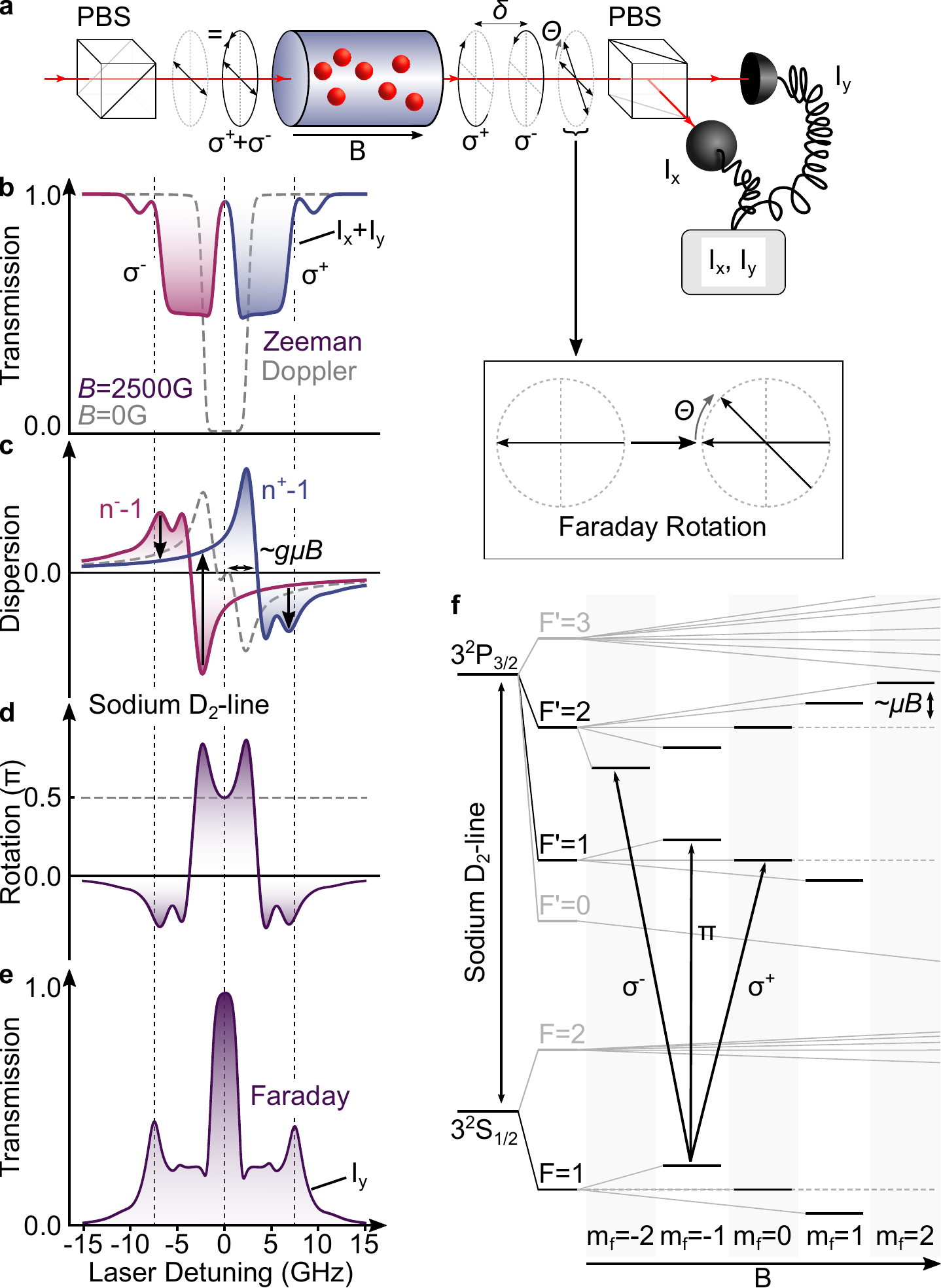}
	\caption{\textbf{General working principle of Faraday filters. a) adapted from~\cite{franke-arnold_jpb_2001}}} %Setup of a Faraday filter. \textbf{b} The applied magnetic field $B$ splits the Doppler spectrum. The direction of the frequency shift of the Zeeman spectrum correlates with the circular polarization $\sigma^{+}$ and $\sigma^{-}$ of the linear polarized light. \textbf{c} The dispersion relation of the absorption feature extracted by the Kramers-Kronig relation for each circular component of the light. Both circular components experience different refractive indices $n^{-}$ and $n^{+}$ in the vapor cell. The splitting between both dispersion relations is proportional to the magnetic field $B$ and hence the Zeeman splitting. \textbf{d, e} The rotation of the light is proportional to the difference of the refractive indices $n^{-}$ and $n^{+}$. For a given cell length and given detuning, the total rotation angle of the linear polarized light varies, changing the amount of transmitted light through the crossed polarizers configuration. \textbf{f} Zeeman level splitting due to a magnetic field and some allowed linear $\pi$ and circular $\sigma$ transitions.}
	\label{fig:03}
\end{figure*}

Figure~\ref{fig:03}b shows the Doppler broadened absorption spectrum for sodium at 153~\degree C. An external magnetic field (here 2500~G) splits the hyperfine states into Zeeman components, which shift towards higher and lower frequencies. As mentioned earlier, the transitions between ground and excited Zeeman states are polarization dependent. The orientation of the atomic dipole vs. the amplitude of the linear polarized excitation light is orthogonal and therefore suppresses any interatomic $\pi$ transitions. However, excitations involving circularly polarized light are allowed. As linear polarized light is the superposition of the two circular components $\pi = \sigma^{+} + \sigma^{-}$, it can drive the transitions which require circular polarizations. The even split between the circular components explains why the two broad Zeeman features in figure~\ref{fig:03}b) drop to roughly 50\% transmission when linear polarized light traverses the cell.

%% the dispersion 
Absorption features relate to the light's dispersion given by the Kramers-Kronig relation, which brings the refractive index into play (as shown in figure~\ref{fig:03}c). When near resonant linear polarized light traverses the vapor cell, each circular light component ($\sigma^{+}$ and $\sigma^{-}$) experiences a different refractive index ($n^{+}$ and $n^{-}$). The dispersion relations of these two circular components show a frequency gap proportional to the Zeeman splitting $\sim \mu B$. The difference between the two dispersion curves represents the optical rotation in the cell, as figure~\ref{fig:03}d visualizes. The resulting sign of the difference in the dispersion curve indicates the direction of the polarization rotation. Of course, the sign changes when reversing the magnetic field direction.

%% the optimal rotation
The highest rotation angle does not automatically mean the highest transmission through the second polarizer. The effective rotation depends on the magnetic field strength and the length of the vapor cell. The transmission through the second polarizer is maximum when the polarization rotation is $\pi/2$ (or integer multiples of this value).

%% a special case - zero rotation
The Faraday transmission spectrum in figure~\ref{fig:03}e) shows some interesting aspects. We first focus on the point where the effective rotation of the polarized light seems to be zero, see zero crossing in figure~\ref{fig:03}d). One might think that when no polarization rotation occurs, the second polarizer blocks all the light. However, we can see a transmission of roughly 25\% in figure~\ref{fig:03}e). What is happening here? To answer this, we first have to clarify the structure of the Zeeman splitting. The two Zeeman features gather either  $\sigma^{+}$ transitions for positive detunings or $\sigma^{-}$ transitions for negative detunings. For a detuning around -3~GHz, there would not be a net rotation because the $\sigma^{-}$ light gets completely absorbed. The light then consists solely of $\sigma^{+}$ light, which does not interact with the atomic medium since the Zeeman levels talking to $\sigma^{+}$ light shifted away to the positive side of the detuning. Therefore, the $\sigma^{+}$ portion of the light reaches the second polarizer unperturbed. Due to its circular nature, only half of it traverses the linear polarizer. As initially explained, the initial linear polarized light consists of an equal ratio of both circular polarizations. Subsequently, only a quarter of the initial light passes the second polarizer. That is why the transmission only shows 25\% at -3~GHz detuning in the Faraday transmission spectrum and also explains the $1/4$ prefactor in equation~\ref{eq:transmission}.

%% a special case - increased wings
Another interesting point is the slightly increased transmission features at the side wings of the Faraday transmission spectrum, marked by the dotted line (around $\pm$7~GHz). At those points, the Zeeman spectrum in~\ref{fig:03} a) shows that neither of the $\sigma^{+}$ or $\sigma^{-}$ components are fully absorbed (only roughly half of it). That means only half of the $\sigma^{\pm}$ light interacts with the Zeeman states. The other half of the light which hits the second polarizes is still the initial $\sigma^{\pm}$ light and undergoes a polarization rotation, following equation~\ref{eq:rot}. That portion of light tops up the initial 25\% of the transmitted light by an additional 20\%.

\section{Experimental Guide}

\subsection{The Doppler Configuration}
%% intro, why rubidium
We now continue with the experimental implementation of an atomic vapor filter. In the previous section, we discussed filters based on sodium. While the sodium filter is an excellent example to introduce the fundamental working principles, filters based on rubidium and cesium are significantly easier to implement in the lab. The reason for this is the higher vapor density at low temperatures for higher-ordered alkali metals. Therefore, these alkali metals allow for building a more dominant absorbing filter at convenient temperatures. For the rest of this manuscript, we exemplary discuss the implementation of a rubidium Doppler and Faraday filter. Compared to other alkali atoms, rubidium allows the use of lower temperatures and magnetic fields to build an efficient atomic vapor filter. Moreover, rubidium resonant lasers are common these days. Dick and Shay reported such a rubidium-based filter in 1991~\cite{dick_ol_1991}.

%% how simple this all is, what comes next
Experiments in the Doppler configuration represent one of the simplest setups for atomic spectroscopy equipment-wise. Such experiments only need three parts: the vapor cell, the light source, and the detector. This straightforward setting is part of the following paragraphs, in which we also discuss further experimental subtleties to circumvent complications.

%% laser selection
Laser diodes are massproduced in spectral regions around 785~nm. These diodes can be pre-selected and fine-tuned to the spectral region of interest. For atomic rubidium, these are the typical D-line transitions around 794.978~nm and 780.242~nm. We selected the rubidium D$_2$-line since it is closer to the characteristic operation wavelength of the laser diodes. Another advantage of the D$_2$-line is its absorption strength, which is compared to the D$_1$-line twice as high due to the higher absorption oscillator strength~\cite{steck_script_2003}. The atomic spectra in this manuscript are recorded far below the optical saturation of the atoms. The probing light intensity has to be so weak that no multiple excitations or higher non-linear effects can occur. A few $\mu$W of optical laser power are sufficient for a beam radius of roughly 3.5~mm.

%% laser scanning
The laser source continuously scans across the rubidium D$_2$ line in a sawtooth shape with 10~Hz. The width of the absorption spectrum varies with the temperature and magnetic field settings. In our case, the width of the spectrum and, therefore, the mode-hop-free tuning range comes down to 10-15~GHz. This scanning range is hard to achieve with a bare laser diode. It is advantageous to use an external cavity diode laser in the Ricci-H\"ansch design~\cite{ricci_opcom_1995}. By carefully adjusting the operating temperature and the current feed-forward, it is simple to achieve a mode-hop-free tuning range exceeding 30~GHz. This type of laser can be either self-build or commercially bought. The frequency scanning speed is 11~Hz.

\begin{figure*}[t]
	\includegraphics[width=\textwidth]{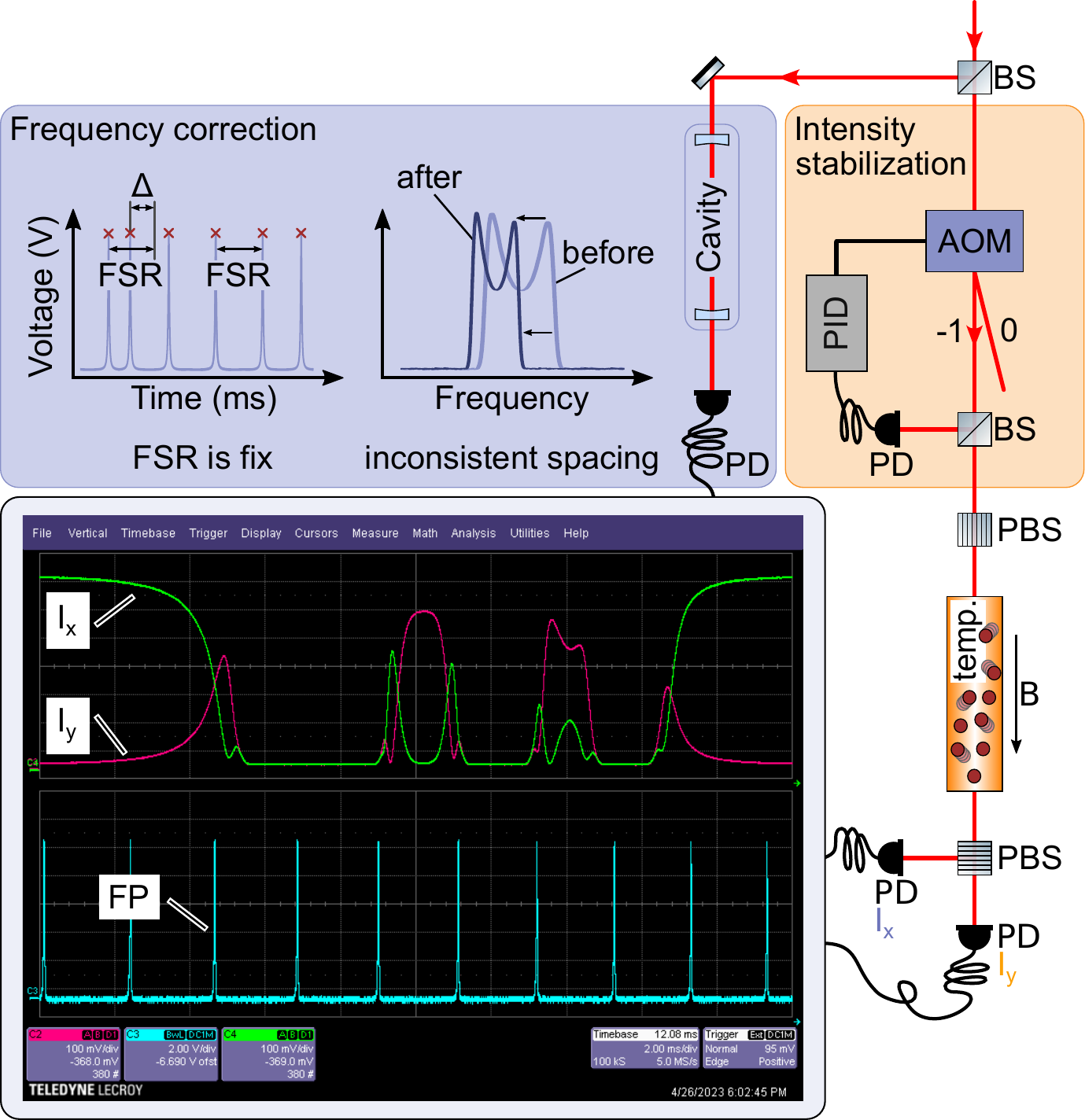}
	\caption{\textbf{Setup and Aquisition:} Setup and Data Acquisition: Basic setup with intensity stabilization by a feedback loop. The photodiode (PD) records the background intensity and corrects via a PID controller the RF amplitude of an acousto-optical modulator with a 350~MHz input. For the excitation beam, we utilize the -1 order. The free spectral range (FSR) of a cavity is the reference to even out the hysteresis of the laser frequency scan. Two photodiodes record the spectra for both polarization, $I_x$ and $I_y$. $I_y$ becomes important for the Faraday filter. $I_x + I_y$ reproduces the Doppler spectrum.} 
	\label{fig:04}
\end{figure*}

%% laser problems: linearization & power 
The robustness and easy handling of such tunable lasers does not automatically shoot for a successful recording of an atomic spectrum. There are two problems with such laser systems: the non-linearity of the frequency scan and power fluctuations due to the change of frequency and polarization drifts. It is required to record a ``ruler'' along with the spectrum. We decided to build a free space Fabry-P\'erot etalon and record its transmission spectrum along with the atomic absorption spectrum. To linearize the absorption spectrum, we apply a peak finding routine to the recorded cavity peaks and fit an Airy-transmission function to the peaks' displacement. D.\ Pizzey and coworkers~\cite{pizzey_njp_2022} published a more extensive explanation of this method. The second problem is power fluctuation induced by the frequency change of the scanning laser or due to polarization drifts in fibers. One reason why polarization drifts occur is when the adjustment of the optical axis set by the fiber dock is slightly off. In this case, even the slightest temperature change impacts the intensity of the light coming out of the fiber. A polarimeter helps to adjust the fiber dock and minimize those intensity fluctuations. There are several methods to even out the intensity fluctuations caused by the laser frequency scan. One is to even out the intensity fluctuations of the spectrum by the evaluation algorithm, explained in~\cite{pizzey_njp_2022}. Our approach was to use a feedback loop to an intensity controller of an acousto-optical modulator. Figure~\ref{fig:04} shows a general scheme for the data acquisition and evaluation.

%% to the vapor cell
The continuously scanning and stabilized laser light then passes the vapor cell. There are several companies and optic suppliers which manufacture such glass cells. For simplicity, we suggest using a 75~mm to 100~mm long cylindrical cell with an excess amount of rubidium of natural abundance. A simple cell out of borosilicate glass and no buffer gas (vacuum) is sufficient. Under ambient conditions (around 20~\degree C), the vapor density is already sufficiently high when using the rubidium and a short vapor cell (around 50~mm). The highest absorbance under such conditions could achieve 60\% for selected atomic transitions. At elevated temperatures around 50\degree C, a 50~mm long vapor cell shows an absorbance of around 90-99\%. Such conditions make the observations of an absorption spectrum easy. We settle for a 100~mm long evacuated rubidium vapor cell. For convenient laser alignment, one can observe the scattered light of the cell with an infrared viewer. The cell lights up when the laser scans over an atomic resonance.

%% heating the cell, intro
Such simple experiments already work under ambient conditions. We already discussed the interplay of temperature and atom density and its importance to optimal settings of the atomic filter. Those factors can be controlled by heating the vapor cell to its desired temperature. Here, there are a few things one should pay attention to. One of them is the heat gradient inside the vapor cell. Most atoms gather at the coldest section of the vapor cell and get stuck at the cell wall. Therefore, it is good practice to heat the cell homogeneously but keep the vapor cell's windows slightly warmer. Otherwise, the rubidium starts to condense on the cell windows, which leads to variations of the beam shape and, in the worst case, blocks the laser completely.

%% heating, how it is done here:
For our experiments, we have chosen to directly heat the optical windows of the cell with electric heat cartridges. Copper blocks, which surround both ends of the vapor cell, heat the area around the window in a homogenous manner. These blocks also feature sockets for temperature sensors. A second pair of windows attached directly to the copper block prevents unwanted airflow and cooling effects of the cell windows. The rest of the cell, including the cell's filling stem, is thermally isolated against the environment. It can take up to a couple of hours to reach thermal equilibrium, which implies a small temperature gradient across the cell. When working with higher temperatures, the air convection around the outer windows might lead to intensity fluctuations. Therefore, we decided to build an even larger enclosure around the vapor cell assembly, which is required to record accurate spectra.

%% heating magnetic fields
The electric heaters can induce an unwanted magnetic field in the vapor cell. Therefore, we perform heating with an alternating current. We decided to use a simple heater with AC-heat cartridges (230~V), tens of watts of electrical power, and a control transformer. Another possibility to reduce the influence of magnetic fields is thick film resistors (Vishay), which operate up to 100~\degree C. A control loop, ideally connected to a PID controller (in our case, Model CN8EPT-330-EIP, Omega), performs a continuous temperature readout and stabilizes it.

%% light in the cell to the diode
The laser light traverses the atomic vapor cell through its optical windows. The wider the beam waist is, the more atoms get excited by the resonant light and contribute to the measured spectrum. The power of the laser should not saturate the sample and should be below the saturation intensity $I_{\rm sat} \approx $3.6~mW/cm$^2$. After the cell, a large-area-amplified silicon photodiode records the transmission. No special modifications of the diode in terms of noise properties or speed is required. Any standard model photodetector from any optics supplier is sufficient. Depending on the beam size and intensity distribution of the beam profile, it might be worthwhile to access different gain settings on the photodiode. Our photodiodes feature different gain settings up to 70~dB (our gain setting was 40~dB).

%% data recording
An oscilloscope records the absorption spectrum. We record three channels simultaneously: the absorption spectrum, the background fluctuations, and the Fabry-P\'erot's transmission. The trigger signal comes from the laser when each frequency scanning loop restarts. We average over eight recordings to reduce the noise level of the absorption spectra. Figure~\ref{fig:04} shows a screenshot of the oscilloscope and its settings. The laser intensity and the spacing of the Fabry-P\'erot comb-like peaks change in the course of the recording. We correct for non-linear frequencies and intensity fluctuations by introducing the above-mentioned schemes and demonstrated in figure~\ref{fig:04}.

\subsection{Observations - Doppler Filter}
%% a first spectrum
Rubidium consists of two different isotopes ($^{85}$Rb and $^{87}$Rb). Due to the ground state splitting (known as the ``clock-transition''), both isotopes have two spectral components. That is why we observe the transitions of one isotope on the inner two dips ($^{85}$Rb) and the other on the outer two dips ($^{87}$Rb). Those four absorption dips inherit three hyperfine transitions for the rubidium D$_2$ line. There are indistinguishable due to the weak splitting of the excited state, which is on the order of hundreds of MHz, compared to the splitting of the two ground states, which is 6.8~GHz for $^{87}$Rb and 3.0~GHz for $^{85}$Rb. The origin of this splitting lies in the addition of quantum numbers of each transition.

%% changing the spectrum
For our rubidium vapor cell of 100~mm length, a temperature around 50~\degree C is sufficient to reach an absorbance above 90\%. At elevated temperatures of 80~\degree C, we observe that the two isotopes $^{87}$Rb and $^{85}$Rb can not be resolved anymore in the left part of the spectrum on a linear scale. For temperatures above 145~\degree C, the spectral components merge into a 10~GHz wide single absorption feature. In comparison to the transmission spectra of the rubidium $D_1$ line, the temperatures have to be increased by roughly 10\degree C to end up in similar situations.

%% fitting the spectra
As outlined above, we even out the spectra for the fluctuating changing light intensity and the non-linear frequency scan range. As the recording is straightforward, we assume the spectra to be theoretically perfectly approximated. A convenient tool for this is the simplified version of the atomic line spectra in the Wolfram Research demonstrations project~\footnote{\url{https://demonstrations.wolfram.com/SpectraOfTheDLinesOfAlkaliVapors/}} or the Python package ElecSus~\footnote{\url{https://github.com/durham-qlm/ElecSus}}. With such tools, it is easy to extract the cell temperature of a recorded spectrum. In our experiment, we observe roughly an 10\degree C deviation from the temperature of the copper blocks and the atomic vapor temperature determined by the fit. Such deviations are often reported in the literature, also for Faraday filters (e.g.~\cite{dick_ol_1991}).

%% Spectral problems
The linearization of the absorption spectrum shows good results when we compare the theoretical model with the data. The calculated residuals are all in the lower percentage range when we focus on the position and width of the absorption spectrum. The steeper slopes are naturally more sensitive and show high deviations from the theory. We account for those deviations due to residual noise and averaging of the spectra.

\begin{figure*}[t]
	\centering
	\includegraphics[width=\textwidth]{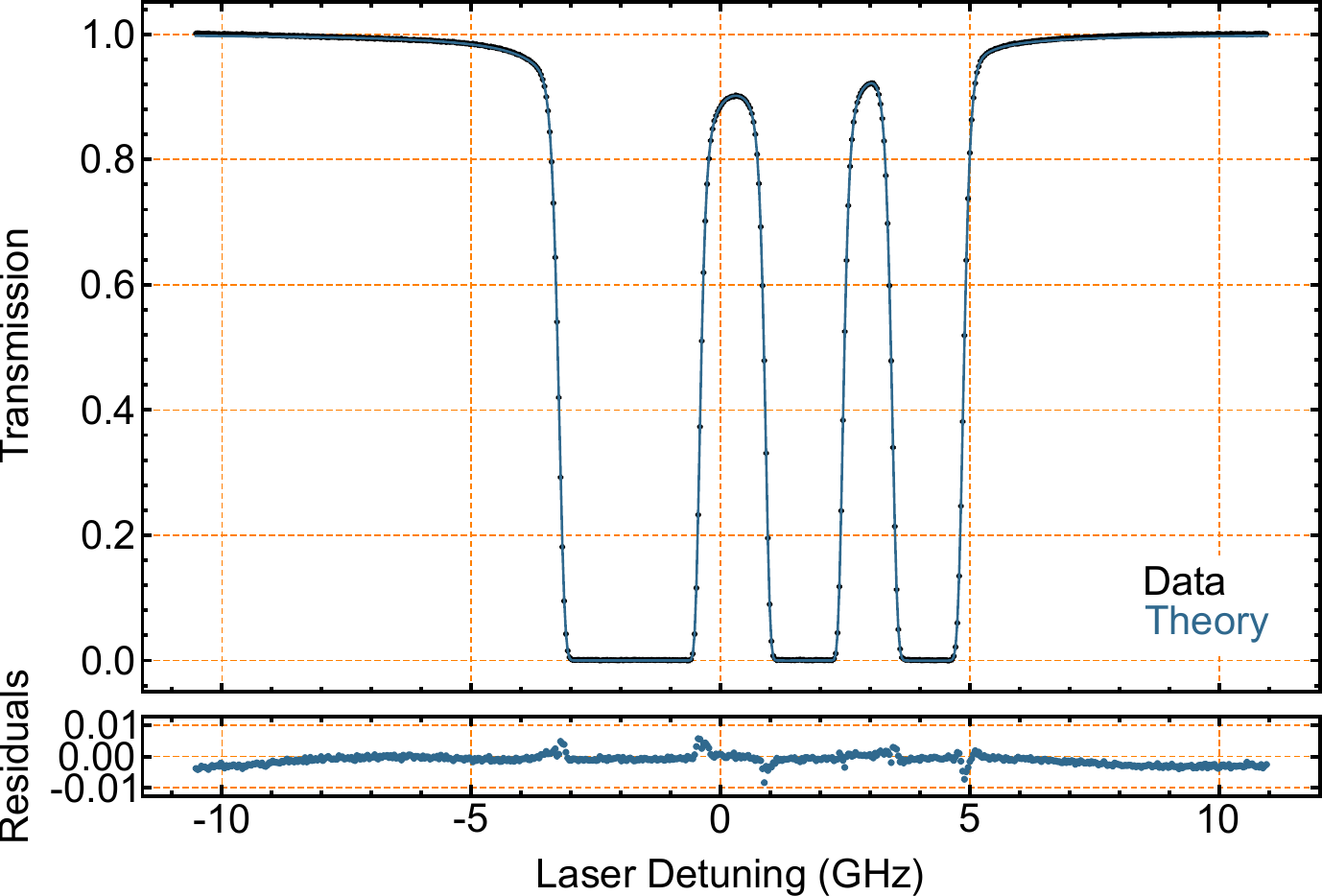}
	\caption{\textbf{Spectrum of a Doppler filter on the rubidium D$_2$ line:} The frequency range of the laser spans roughly 20~GHz wide. For the 100~mm cell and a laser power of 15~$\mu$W, the fitted model suggests a temperature of $T=79.11 \pm 0.02\degree$C. The calculated residuals go as low as $\pm$0.2\% for the plateaus and are below $\pm$1\% at the steeper edges.}
\end{figure*}

\subsection{Introducing a Magnetig Field: Towards the Faraday Filter}
%% introduce magnetic field
Up to this point, the focus of this section was on the Doppler spectrum. We have not strictly defined the incident polarization of the excitation light, nor did we account for residual magnetic fields. The incident light polarization was of no urgent importance for the Doppler case. In the absence of a magnetic field, no further splitting of the hyperfine states occurs, and therefore, the initial energy states are all degenerated. Also, the earth's magnetic field does not have a significant impact on the spectrum. This makes the absorption features in the Doppler configuration polarization independent. The vapor cell sits in a solenoid with around 3000 windings of 0.75~mm enameled copper wire. This coil is able to provide a field of approximately 500~G. This magnetic field is more than sufficient for rubidium (or cesium) with a cell length of 10~mm or larger. In comparison, sodium requires magnetic fields up to 2000-2500~G for an optimal working Faraday filter caused by the lower vapor density. In the past, experiments with permanent magnets were performed~\cite{widmann_arxiv_2015, portalupi_nc_2016, pizzey_rsi_2021}. Such attempts are limited due to two factors: a) the magnetic field is limited for large cells, and b) since the cell is heated, one has to account for the magnet's Curie temperature. A good amount of the heat does not reach the atomic vapor cell but gets distributed over the entire experimental configuration, including the magnet.

\subsection{The Laser}
%% laser scan range
In the experiment, the laser scans across the rubidium D$_2$ line. Compared to the Doppler spectrum, the scan range should be larger for the Faraday filter. The wings of the Faraday spectrum show a more shallow slope, unlike the sharp Doppler-limited components. Ideally, the mode-hope-free detuning range of the laser exceeds 16~GHz for the rubidium D$_2$ Faraday filter.

\subsection{The Influence of Polarization}
%% incident polarizer
To define the polarization of the light before the vapor cell, we apply a polarizing beam splitter on a fixed optical mount. That way, we ensure that only linear polarized light enters the vapor cell. 

%% outgoing polarizer
It is convenient to monitor both linear components of the excitation light behind the vapor cell. Therefore, a second polarizing beam splitter is placed right behind the vapor cell. Two photodiodes record both outputs of the polarized beam splitter simultaneously. That way, we can reconstruct the Doppler signal by adding them together. The alignment of both polarizers has to be such that one photodiode does not detect any light without an external magnetic field (polarizers cross out the light). Ideally, one performs this mechanical alignment with far-off resonant light so that small residual magnetic fields do not have any influence. The crossed-polarizers are also an indicator to catch and prevent saturation effects. If the laser intensity is too high, one observes some polarization changes close to an atomic transition. In this case, unexpected light breaks through to the photodiode, characterized by small revival peaks in the absorption spectrum. When reducing the intensity below saturation, the revival peaks vanish. Ideally, the extinction ratio of the two polarizers reaches up to five orders of magnitude. Often, this is limited by some residual birefringence of the vapor cell's windows. The extinction ratio is also commonly diminished by a large beam due to lateral variations of the window's retardance.

%% polarization neutral filter
Sometimes it is worthwhile to exchange the polarized beam splitter with a calcite beam displacer. Those elements split the incident polarization into two different linear components. Behind the cell, the resulting spectra can be analyzed independently of the incident polarization of light. This is an interesting option for experiments in quantum optics where e.g. single photons are encoded on a polarization basis~\cite{bennett_2014}.

%% compensation for the windows
The theory section already pointed out that the polarization of the incident light is a critical point for the Faraday filter. Little variations of the polarization result in changes in the recorded spectrum. Such little variations can occur due to the residual birefringence of the vapor cell windows. A $\lambda/2$ and a $\lambda/4$ waveplate in the right setting corrects for the birefringence effects. This becomes important when the spectrum should match the theoretical predictions.

\begin{figure*}[t]
        \centering
        \includegraphics[width=\textwidth]{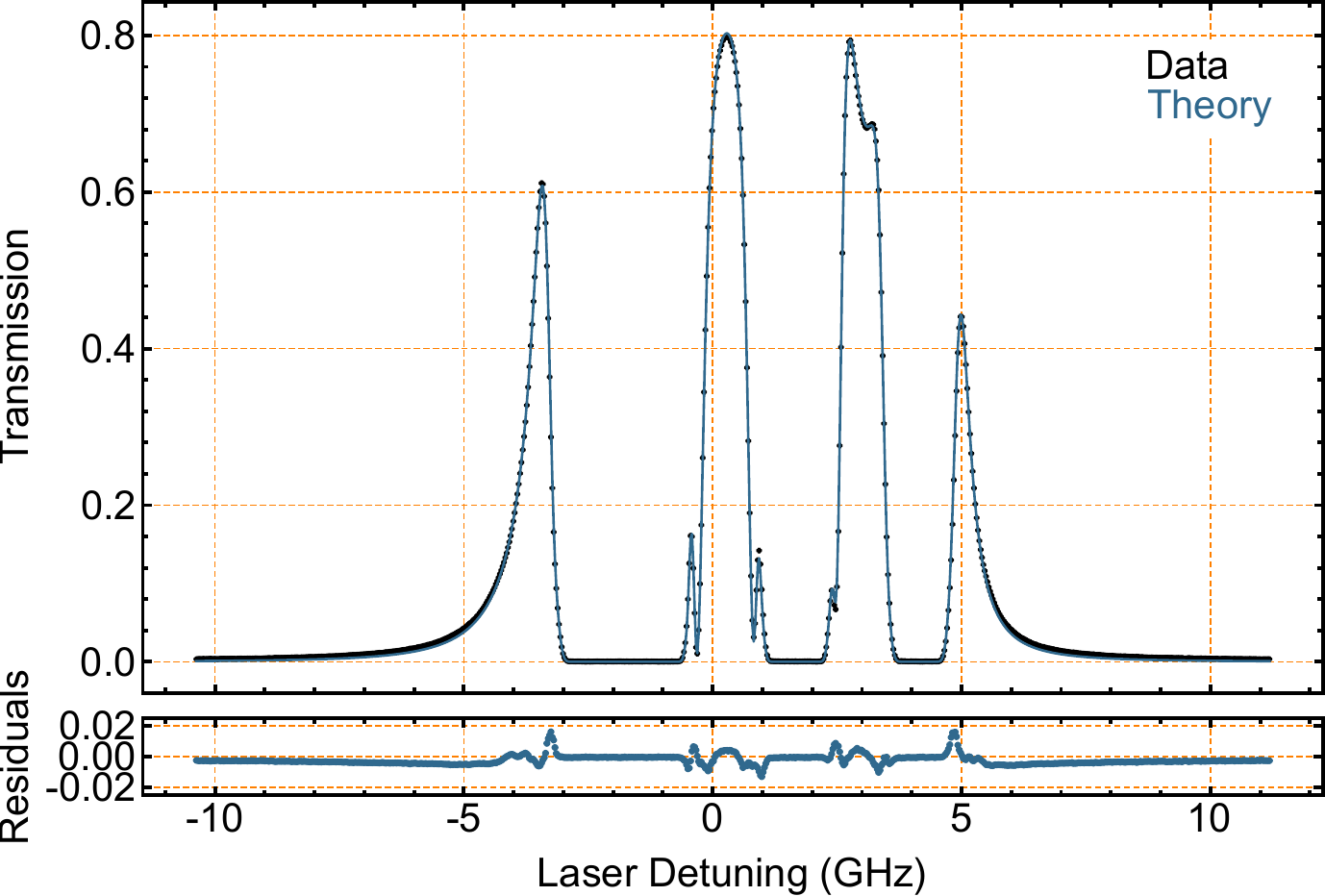}
        \caption{\textbf{Spectrum of a Faraday filter on the rubidium D$_2$ line:} The frequency range of the laser spans roughly 20~GHz wide. For the 100~mm cell and laser power of 7$\mu$W, the fitted model suggests a temperature of $T=79.12 \pm 0.03\degree$C and a magnetic field of $B = 43.85 \pm 0.3$~G. The calculated residuals go as low as $\pm$0.2\% for the plateaus and are around $\pm$2\% on the steeper edges.}
\end{figure*}

\subsection{Observations - The Faraday Filter}
%% What's observed
The measured Faraday spectrum corresponds closely to the observed spectrum by Dick and Shay~\cite{dick_ol_1991}. Unlike the Doppler absorption spectrum, the Faraday spectrum consists of four different components. These are simply the rims of the Doppler spectrum. The closer the spectral components are to the spectral line, the higher the optical rotation. Note that both, the Doppler and Faraday spectrum, were recorded on two different days and can show minor deviations for the set temperature.

\section{Conclusion}
In conclusion, we have outlined a hands-on guide on how to understand and apply the foundation of optical filters based on the interaction of light and hot atomic vapors. The focus was on the Doppler and the Faraday filter and highlighted their narrow-banded character, the implementation of such filters, and a guide on how to build such a setup that follows theoretical predictions closely. 

Both filters are easy to implement. Most notably, the Doppler filter is considered a fundamental atomic physics experiment. This interlinks our article to the more fundamental view of Pizzey and coworkers~\cite{pizzey_njp_2022}.

Our view on atomic filters was limited to a small number of magneto-optical effects. A large collection of these effects has been outlined before~\cite{budker_rmp_2002}. Still, we hope that the experimental description facilitates more experiments that utilize atomic vapor filters in a larger variety. We see a large number of experimental implementations in quantum optics reaching up to microbiology, where such filters might lead to a crucial advantage like increasing the photon detection efficiency by a margin~\cite{uhland_epj_2015}, such that the laser light intensity can be decreased and therefore makes biomatter longer sustainable. Space weather applications, in particular the prediction of solar flares, use atomic vapor filters in telescopes to measure the magnetic and Doppler fields of such events~\cite{korsos_apj_2020, erdelyi_swsc_2022}. Another interesting application could be the analysis of Stokes parameters, which can be used to build an effective polarization measurement tool~\cite{weller_jpb_2012}.

\section{Acknowledgments}
The project was funded by the Deutsche Forschungsgemeinschaft with the project GE 2737/5-1 and the Bundesministerium f\"ur Bildung und Forschung (13N15972).

\section*{References}
\bibliographystyle{unsrt}

\newpage
\section{Appendix A -- Faraday Filter Setup}
\begin{figure}[ht]
	\centering
	\includegraphics[width=\textwidth]{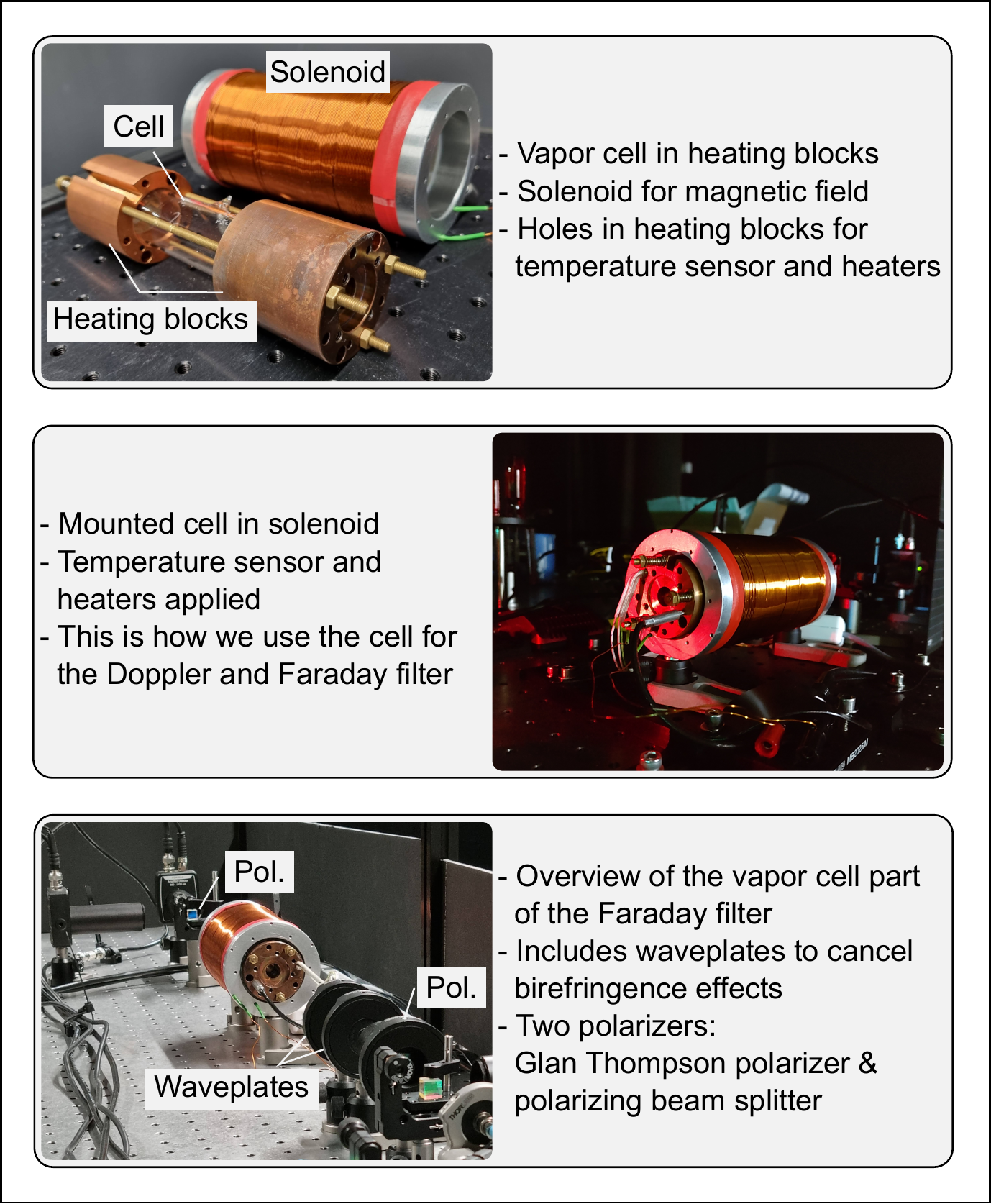}
	\caption{Pictures from the setup}
        %\label{fig:07}
\end{figure}

\newpage
\section{Appendix B -- Schematics Magnetic Coil}
\begin{figure}[ht]
	\centering
        \includegraphics[width=0.95\textwidth]{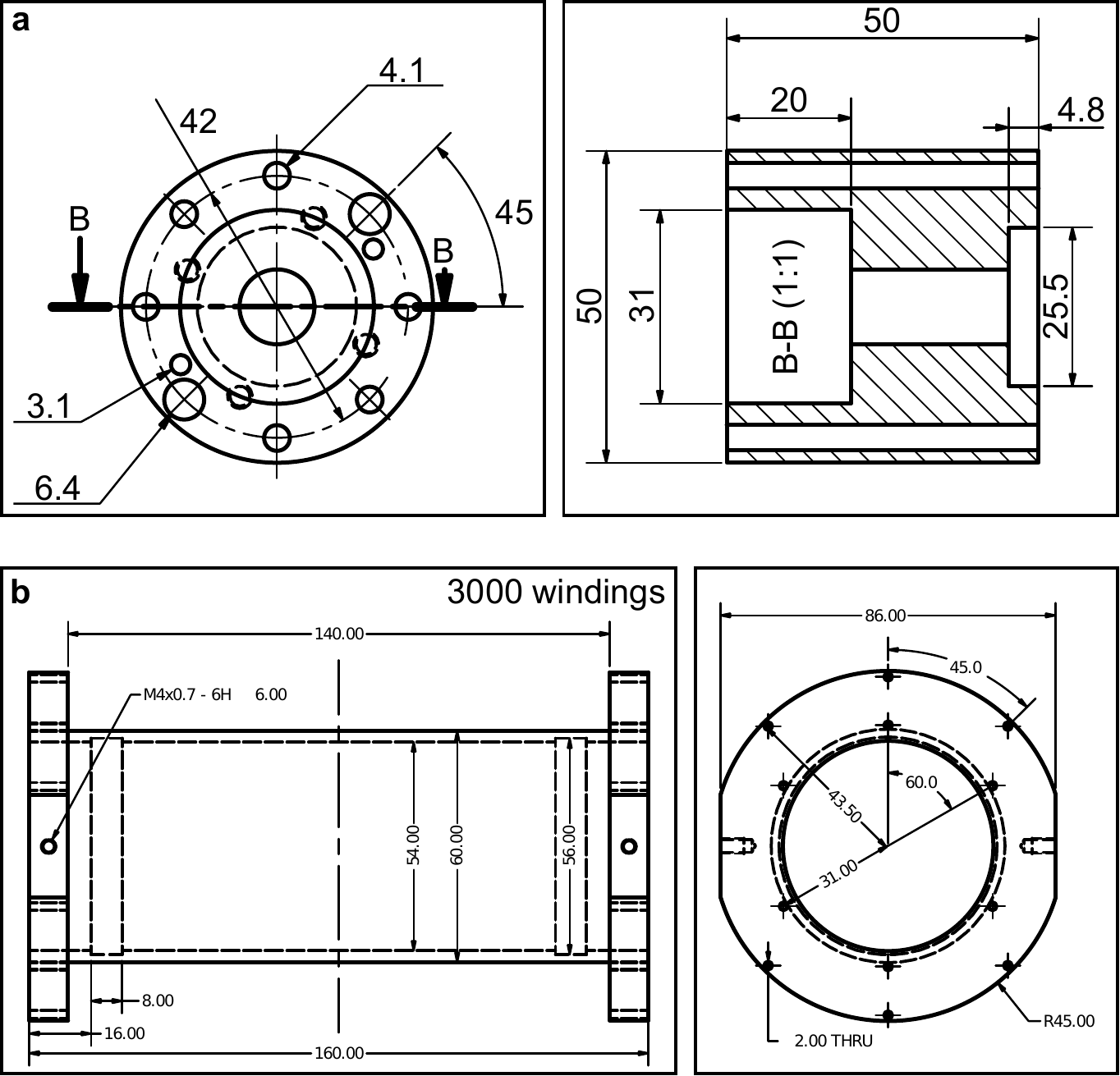}
	\caption{The magnetic coil in use has 3000 windings and 0.75~mm (diameter) enameled copper wire. All units are milimeter.}
        %\label{fig:07}
\end{figure}

\newpage
\section{Appendix C -- Finding the Optimal Settings for T, B, and L}
\begin{figure}[ht]
	\centering
        \includegraphics[width=0.95\textwidth]{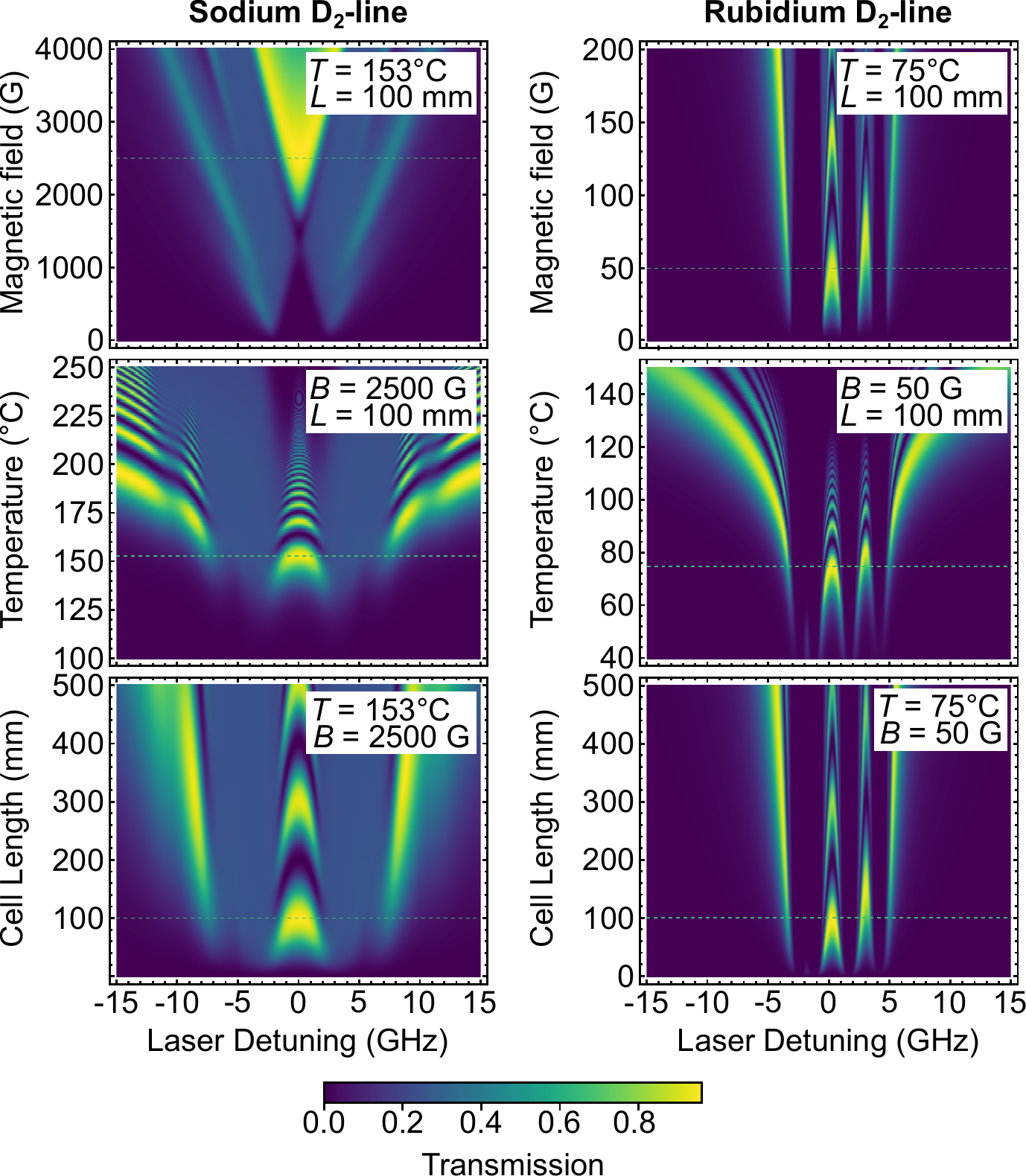}
	\caption{Simulation of the transmission for the magnetic field $B$, the temperature $T$ and the cell length $L$ for different laser detunigns. This was done for sodium and rubidium around the D$_2$ line.}
        \label{fig:ap:c}
\end{figure}

\noindent
The functionality of the Faraday can vary depending on someone's needs. Therefore, one should decide to suppress or allow specific spectral frequencies. One approach to measuring the efficiency of the Faraday filter is by maximizing the transmission for a given laser frequency. The optimal point is when: 1) the Doppler absorption is effectively zero, and 2) the rotation for linearly polarized light is $\pi/2$. Then, the transmission of the filter is maximal. That requires an optimal selection of the three parameters: the magnetic field strength $B$, the vapor temperature $T$, and the filter length $L$. Figure~\ref{fig:ap:c} shows a simulation of where to find the maximum transition points for a set of given parameters (for sodium and rubidium around the D$_2$ line). The dotted line indicates the ideal setting for a maximum (close to a $\pi/2$ rotation of the polarization) at zero detuning.  

\end{document}